\documentstyle[12pt]{article}
\font\titlefont=cmbx10 scaled \magstep3
\begin{document}

\begin{center} {\titlefont Limitations on quantum
information from\\ black hole physics}\footnote{Lecture
delivered at the XXV School of Theoretical Physics, 
Ustro\'n, Poland, 10-16 Sep. 2001.} \\
 
\vskip .3in Jacob D. BEKENSTEIN
\vskip .1in The Racah Institute of Physics, Hebrew
University of Jerusalem\\ Givat Ram, Jerusalem 91904,
Israel\\
\end{center}

\vskip .2in

\begin{abstract} After reviewing the relation of entropy to
information, I derive the entropy bound as applied to
bounded weakly gravitating systems, and review the bound's 
applications to cosmology as well as its extensions to
higher dimensions.  I then discuss why black holes  behave
as 1-D objects when emitting entropy, which suggests that
a  black hole swallows information at a rate restricted by
the one-channel information capacity.  I discuss
fundamental limitations on  the information borne by signal
pulses in curved spacetime, from  which I verify the
mentioned bound on the rate of information  disposal by a 
black hole.
\end{abstract}
\vskip .1in PACS categories: 04.70.-s, 89.70.+c, 03.67.Hk, 
04.70.Dy, 65.50.+m
  
\baselineskip=14pt

\section{Introduction} Were one looking for a logo to
symbolize the field of gravity physics in the last decade,
none would be more apt than 't Hooft's holographic bound
\cite{thooft}: the entropy $S$ (or information---see below)
that can be contained in a physical  system is bounded in
terms of the area
$A$ of a surface enclosing it:
\begin{equation}  S\leq A(4\hbar)^{-1}
\label{holographic} 
\end{equation}  (I assume units with $G=c=1$).   Where does
this comes from ? According to Susskind
 \cite{susskind}, the holographic bound is required  by the
generalized second law (GSL) \cite{bek72} applied to the
total collapse of a physical system into a black hole of
its own making. Granted that many systems do not like to
collapse spontaneously, so that this argument is not a
general proof of the principle \cite{wald}, a similar
argument of wider applicability can be given for quiescent
systems by considering either infall of the system into a
large black hole, or a tiny auxiliary black hole which
devours the system
 \cite{bek00}.  The holographic bound as above stated can be
violated by rapidly evolving systems, but with Bousso's
reinterpretation \cite{bousso1} of the meaning of
$A$ in Eq.~(\ref{holographic}), it works in these cases
also.

Interest in the holographic bound, or the more encompassing
holographic principle \cite{thooft}, is mostly for what it
tells us about the structure of physical laws.  However, it
is clear that the holographic bound also serves as the
final arbitrator of the promise of any futuristic
information storage technology. Unfortunately, it is rather
lenient in this respect: it merely requires that a device 
with dimension 1 cm hold no more than some $10^{66}$ bits
of information.  By contrast, all the books in the Library
of Congress hold a paltry $10^{15}$ bits of information,
and no state-of-the-art 1 cm size memory can hold all
that.  So I would like to ask if one can device a tighter
bound on information storage than the holographic one ?  As
I will show, the answer is positive in the form of the
universal entropy bound, a spinoff of black hole physics. 
In light of the explosive development of fast
communications, a further interesting question would be:
what fundamental bounds can be set on the {\it flow\/} of
information ?  I will show here that several new results in
this well developed field can be had by considering black
holes.

\section{Information and Entropy}
\label{sec:I&E}

I want to start by recapitulating how thermodynamic entropy
of a quantum system and the information it can store are
related. In quantum theory a system's {\it state\/},
whether pure or mixed, is described by an hermitian
operator $\rho$ with unit trace. The entropy of a pure
state is zero (because we know what state we are dealing
with).  The thermodynamic entropy of a mixed state is given
by von Neumann's formula $S=-{\rm Tr}\, \rho\ln \rho$
(which will properly give zero if $\rho$ corresponds to a
pure state: ${\rm Tr}\, \rho^2=1$).

Suppose we find the eigenstates of $\rho$ and their
respective eigenvalues, which must obviously sum to unity
because ${\rm Tr}\rho=1$.  According to quantum theory,
each nonvanishing eigenvalue $p_i$ stands for the
probability with which the corresponding eigenstate turns
up in state $\rho$.  Working in the basis furnished by the
eigenstates it is easy to see that $-{\rm Tr}\,
\rho\ln \rho=-\sum_i p_i\ln p_i$.  Now compare this result
with  Shannon's famous 1948 formula
 \cite{shannon} for the peak information capacity---or
information entropy---in bits of a system with
distinguishable states which occur with {\it a priori\/}
probabilities $\{p_i\}$:
\begin{equation} I_{\rm max} = -\sum_i p_i\, \log_2 p_i
\label{S_shannon}
\end{equation}  We obviously have $I_{\rm max}=S\log_2 e$:
generically thermodynamic entropy sets an upper bound on
the information storage capacity of the system, modulo the
factor $\log_2 e$ which converts from natural entropy units
to bits.

The key word in the above identification is
``distinguishable''.  The eigenstates of $\rho$ are all
orthogonal by virtue of its hermitian character, so they
are precisely distinguishable.  Confusion can occur (and
has often) if the distinguishability condition is
compromised.  Consider the four states
$|\uparrow\rangle, |\downarrow\rangle, |\rightarrow\rangle,
|\leftarrow\rangle$ of a spin $1/2$ particle corresponding
to ``up'' and ``down'' spin with respect to spin components
$s_z$ and
$s_x$, respectively.  If we assign each probability
${\scriptstyle 1\over \scriptstyle 4}$, Shannon's formula
would predict maximum information capacity $I_{\rm
max}=\log_2 4=2$. Obviously here
\begin{equation}
\rho={\scriptstyle 1\over \scriptstyle
4}\left(|\uparrow\rangle\langle
\uparrow|+ |\downarrow\rangle\langle
\downarrow|+|\rightarrow\rangle\langle
\rightarrow|+|\leftarrow\rangle\langle
\leftarrow|\right)
\label{rho}
\end{equation} Recalling that
$|\rightarrow\rangle=2^{-1/2}(|\uparrow\rangle+|\downarrow\rangle)$
while
$|\rightarrow\rangle=2^{-1/2}(|\uparrow\rangle-|\downarrow\rangle)$
and that $|\uparrow\rangle$ and $|\downarrow\rangle$ are
orthonormal states,  we easily work out that
$\rho={\scriptstyle 1\over
\scriptstyle 2}\left(|\uparrow\rangle\langle
\uparrow|+ |\downarrow\rangle\langle
\downarrow|\right)$, which form tells us immediately that
$\rho$'s eigenvalues are
$\{{\scriptstyle 1\over \scriptstyle 2},{\scriptstyle 1\over
\scriptstyle 2}\}$ and its eigenstates
$\{|\uparrow\rangle,|\downarrow\rangle\}$.  Calculating in
the basis
$\{|\uparrow\rangle,|\downarrow\rangle\}$ gives $S=\ln 2$. 
We thus find that $I_{\rm max}=2S\log_2 e$.

Does the above mean the spin system can store more
information than allowed by the thermodynamic entropy ? 
Not at all !  The four states used are not independent, and
so not mutually orthogonal.  According to quantum theory
nonorthogonal states are not fully distinguishable
experimentally \cite{peres}.  Thus when reading out
information encoded in our four states, errors can occur,
and this ``noise'' reduces the true information extractable
below Shannon's formal $I_{\rm max}$.  According to a
fundamental theorem by Holevo \cite{holevo},  the
information that can actually be read out of a system is
bounded from above by $S\log_2 e $, as claimed above.

Some confusion can still arise if we ignore the fact that
von Neumann's $S$ can have various values depending on the
level of structure at which it is computed. The chemist,
for instance, determines the entropy $S$ of a piece of iron
by methods that reach down to the atomic level; for him the
states $\{i\}$ are atomic states.  The communication
engineer, by contrast, is interested in storing information
in the magnetic domains of the iron in a magnetic tape.  He
groups a multitude of atomic states into domain states, and
the $S$ computed from the latter is definitely smaller than
the chemist's $S$.  There is no  contradiction here; one
must simply specify at which level $S$ or the corresponding
information capacity are calculated. Obviously, the deeper
we go into the system's structure, , the higher the entropy
and information capacity.  In what follows I shall be
interested in the entropy (information capacity)
$S_{\rm X}$ calculated at level X, the deepest level of
structure (the level of lepton, quark and gluon degrees of
freedom).  The corresponding
$S_{\rm X}\log_2 e$ bounds from above the information
capacity of material media accessible with any achievable
technology.

\section{Poor Man's Road to the Universal Entropy Bound}
\label{sec"poor_man}

Consider the following {\it gedanken\/} experiment
\cite{MG7}. Drop a physical system ${\cal U}$ of unknown
construction and constitution having mass-energy $E$ and
radius $R$
 into a Schwarzschild black hole of mass  $M\gg E$ from a 
large distance $d\gg M$ away.  This $d$ is so chosen that
the Hawking radiance carries away energy (as measured at
infinity) equal to $E$ while ${\cal U}$ is falling to the
horizon where it is effectively assimilated by the black
hole.   At the end of the process the black hole is back at
mass $M$ and its entropy has not changed.  Were the emission
reversible, the radiated entropy  would be
$E/T_{\rm H}$ with $T_{\rm H}\equiv \hbar(8\pi M)^{-1}$. 
Irreversibility of radiation and spacetime curvature
conspire to make the entropy emitted a factor $\nu$ larger;
typical values, depending on particle species, are
\cite{page2} $\nu=1.35$--$1.64$.  Thus the overall change
in world entropy is     
\begin{equation}
\delta S_{\rm ext} = \delta S_{\rm rad}-S= \nu E/T_{\rm H} -
S  
\label{netchange} 
\end{equation} 

One can certainly  choose  $M$ larger than
$R$, say, by an order of magnitude so that the system will
fall into the hole without being torn up: $M=\zeta R$ with
$\zeta = {\rm a\ few}$. Thus by the {\it ordinary\/} second
law we obtain the bound
\begin{equation}   S < 8\pi\nu\zeta RE/\hbar   
\label{newbound}   
\end{equation}  Our simple argument here leaves the factor
$\nu\zeta$ somewhat fuzzy; but it is safe to say that
$4\nu\zeta<10^2$.  Thus we have obtained a bound on the
entropy of an arbitrary system ${\cal U}$ in terms of just
its total energy $E$ and radius $R$.  This is the gist of
the universal entropy bound \cite{bek81}.  

Note that we could not derive (\ref{newbound}) by using a
heat reservoir in lieu of a black hole.  A reservoir which
has gained energy $E$ upon ${\cal U}$'s assimilation, and
has returned to its initial energy by radiating, does not
necessarily return to its initial entropy, certainly not
until ${\cal U}$ equilibrates with the rest of the
reservoir.  But a (nonrotating uncharged) black hole whose
mass has not changed overall, retains its original entropy
because that depends only on mass.  In addition, for the
black hole mass and radius are related in a simple way; this
allowed us to replace $T_{\rm H}$ in terms of $R$. By
contrast, for a generic reservoir, size is not simply
related to temperature.

Note also that in our argument ${\cal U}$ is not allowed to
be be strongly gravitating (meaning $R\sim E$) because then
$M$ could not be large compared to $E$ while 
$\zeta$ is of order a few, as I assumed.  We thus have to
assume that $R\gg E$ in addition: the universal entropy
bound applies, {\it a priori\/}, only to weakly gravitating
systems  (but see below).

But what about Hawking radiation pressure.  Is it important
?  Could it blow
${\cal U}$ outwards ?   If we approximate the radiance as
black-body radiance of temperature $T_{\rm H}$ coming from a
sphere of radius $2M$, the energy flux at Schwarzschild
coordinate
$r$ from the hole is 
\begin{equation}  F(r)= {{\cal N}\hbar\over 61,440(\pi M
r)^2},
\label{flux}
\end{equation}  where ${\cal N}$ stands for the effective
number of massless species radiated (photons contribute 1 to
${\cal N}$ and each neutrino species $7/16$).  This estimate
is known to be off by a factor of only a few \cite{page1}. 
This energy (and momentum) flux results in a radiation
pressure force $f_{\rm rad}(r)=\pi R^2 F(r)$ on ${\cal U}$. 
More precisely, species which reflect well off ${\cal U}$
are approximately twice as effective at exerting force as
just stated, while those (neutrinos and gravitons) which go
right through
${\cal U}$ contribute very little; the ${\cal N}$ must thus
be reinterpreted accordingly.  I have ignored relativistic
corrections so that the result, as qualified, is correct
mostly for $r\gg M$.  

Writing the gravitational force on ${\cal U}$ in the 
Newtonian approximation,  $f_{\rm grav}(r)=ME/r^2$, one sees
that
\begin{equation}  {f_{\rm rad}(r)\over f_{\rm grav}(r)}=
{{\cal N}_{\rm eff}\,\hbar R^2\over 61,440  \pi^2 M^3 E}
\label{ratio}
\end{equation}  I have written ${\cal N}_{\rm eff}$ here
because, as mentioned, some species just pass through ${\cal
U}$ without exerting force on it.  In addition, only those
species actually represented in the radiation flowing out
during ${\cal U}$'s infall have a chance to exert forces. 
Now an Hawking quantum bears an energy of order $T_{\rm H}$,
so the number of quanta radiated together with energy $E$ is
approximately  $8\pi M E/\hbar$.  However, for any bound
system 
$\hbar/E < R$ (system larger than its own Compton length),
so by our stipulation that
$M=\zeta R>R$, the number of species can be large compared
to unity.  Since a species can exert pressure only if it is
represented by at least one quantum, one obviously has
${\cal N}_{\rm eff} < 8\pi ME/\hbar$.  Therefore,
\begin{equation}    {f_{\rm rad}(r)\over f_{\rm grav}(r)}<
{R^2\over 7680  \pi M^2}
\ll 1 
\label{newratio}
\end{equation}  Radiation pressure is thus negligible, and
${\cal U}$'s fall is very nearly on a geodesic, at least
until ${\cal U}$ approaches to within a few Schwarzschild
radii.  It is intuitively clear that if $d\gg M$, the last
(relativistic) stage cannot make any difference, and ${\cal
U}$ must plunge to the horizon.

Whether $d\gg M$ as assumed must be checked.  I have taken
$d$ such that the infall time equals the time $t$ for the
hole to radiate energy $E$.  Newtonially  $d\approx 2(t^2
M/\pi^2)^{1/3}$, while Eq.~(\ref{flux}) gives the estimate
$t\approx 5\times 10^4 EM^2\hbar^{-1}{\cal N}^{-1}$ with
${\cal N}$ now the full species number.  From these
equations and $M=\zeta R$ we get that $d\approx 1.2\times
10^3 (\zeta ER/{\cal N}\hbar)^{2/3}M$.  Thus for ${\cal
N}<10^2$ (conservative estimate of {\it our\/} world's
massless particle content), we have $d>57 M\gg M$ for all
systems ${\cal U}$ satisfying our assumption $R>\hbar/E$.

\section{Ramifications of the Universal Entropy Bound}
\label{sec:newbound}
    
In asymptotically flat four-dimensional spacetime
($D=n+1=4$) the holographic bound restricts the entropy of
a finite system ${\cal U}$ with energy $E$ and
circumscribing  radius $R$ ($E$ and $R$ measured in proper
frame) by Eq.~(\ref{holographic}).  For weakly gravitating
systems $(R\ll E)$ the universal entropy bound in its
original version \cite{bek81} is
\begin{equation}  S\leq  2\pi ER/\hbar.
\label{original}
\end{equation}  Typically for laboratory sized systems $E<
10^{-23} R$, while for astronomical systems---barring
neutron stars and black holes--- $E< 10^{-5} R$.  Thus,
with few exceptions, the entropy bound is many orders of
magnitude tighter than the holographic one.  For instance,
it  limits the information capacity  of a 1 cm device made
of ordinary matter to be less than $10^{37}$ bits, which
limit no longer looks unreachable.   

In the original derivation of bound (\ref{original}) I
imagined that ${\cal U}$ is lowered slowly from far away to
the horizon of a stationary black hole, while all the freed
potential energy is allowed to do work on a distant agent (a
Geroch process 
\cite{bek73}).  I then applied the GSL to infer the bound.
This derivation was criticized \cite{UW} for not taking into
account the buoyancy of ${\cal U}$ in the Unruh radiation
surrounding it by virtue of its acceleration.  However, I
have shown \cite{bek94} that correction for
buoyancy---itself an intricate calculation---merely
increases the $2\pi$ coefficient in Eq.~(\ref{original}) by
a tiny amount provided only that one assumes that
$R\geq\hbar/E$, as done in Sec.~\ref{sec"poor_man}.  

The entropy in bounds (\ref{newbound}) or (\ref{original})
refers to the
$S_{\rm X}$ defined in Sec.~\ref{sec:I&E}.  This is because
gravitation plays a crucial role in many generic ways of
deriving the entropy bound \cite{bek81,bek94}.   And
gravitation is unique among the interactions in that it is
aware of all degrees of freedom in its sources (according
to the equivalence principle all energy gravitates). It
would thus be odd if the entropy bound took into account
only entropy corresponding to intermediate degrees of
freedom,  and so ignored energy carrying states at the
deeper levels ?

Of late it has been realized that the entropy bound also
applies in higher dimensions.  For instance, Bousso
\cite{bousso00} has shown, via the Geroch process argument,
that bound (\ref{original}) applies {\it verbatim\/} in
asymptotically flat spacetime with any 
$D=n+1$ dimensions.  Bousso rewrites the bound in terms of
${\cal U}$'s gravitational radius $r_{\rm g}$ as inferred
from the D-dimensional Schwarzschild solution, 
\begin{equation}  r_{\rm g}{}^{n-2}={8\Gamma(n/2)
E\over(n-1)\pi^{n/2-1}\ },
 \label{rg}
\end{equation}  whereby it takes takes the form
\begin{equation}  S\leq {(n-1)\pi^{n/2}\,r_{\rm
g}^{n-2}R\over 4\Gamma(n/2)\,\hbar^{(n-1)/2}}. 
\label{bousso}
\end{equation} According to Bousso this form of the bound
to apply to all systems in D-dimensional de Sitter
spacetime which occupy a small part of the space inside the
cosmological horizon whose radius is denoted by
$R$. 

As mentioned in Sec.~\ref{sec"poor_man}, when we come to
strongly gravitating systems ($E\sim R$), we cannot derive
the bound (\ref{original}) or even the weaker version
(\ref{newbound}) by the methods just expounded.   However,
it so happens that the bound (\ref{original}) is actually
obeyed---and saturated at that---by all
$D=4$ Kerr-Newman black holes provided one interprets $E$
as the black hole's mass and $R$ as the Boyer-Lindquist
coordinate of the horizon,
$r_+$ (see references in \cite{erice}). Further, spherical
{\it black holes\/} in $D>4$ spacetime (for which the
horizon has a $(n-1)$ dimensional ``area'') also obey
(\ref{original}) in asymptotically flat spacetimes, and the
bound (\ref{bousso}) in asymptotically de Sitter
spacetimes.  However, black holes in higher dimensions no
longer saturate these bounds as for $D=4$ \cite{bousso00}.  

The sway of the entropy bound also extends to cosmology. 
For example,  E. Verlinde \cite{verlinde} has shown that
the entropy $S$ of a complete closed Robertson-Walker
universe in $D$ spacetime dimensions whose contents are
described by a conformal field theory (CFT)---the deeper
description of a number of massless fields possibly in
interaction---with large central charge (essentially many
particle species), is subject to the generic bound
\begin{equation}  S\leq {2\pi R\over n\hbar }[E_{\rm
C}(2E-E_{\rm C})]^{1/2},
\label{verlinde_bound}
\end{equation}  where $R$ is the radius of the $S^n$ space,
$E$ the total energy in the fields and $E_{\rm C}$ the
Casimir (vacuum) energy (which shows up because the
cosmological space is compact).   Verlinde points out that
for fixed $E$ the maximum of his bound is $2\pi R
E/(n\hbar)$, which never exceeds the original entropy bound
(\ref{original}); indeed Verlinde adopts $S \leq 2\pi R
E/(n\hbar)$ as the fiducial form of that bound.   A number
of recent papers (see
\cite{erice} for references) have substantiated Verlinde's
bound; they culminate years of efforts by many to make
meaningful statements about the entropy  (and by
implication the maximum information) that can be contained
in a whole universe.  

For {\it strongly\/} gravitating systems in asymptotically
flat spacetime with $D=4$, the holographic bound and the
formal  entropy bound make very similar predictions, but
for $D>4$ the holographic bound is the tighter of the two. 
Unless $D$ is very large,  the entropy bound is the tighter
bound for weakly gravitating systems, such as those we meet
in everyday life.

\section{Black Holes as Information Pipes}
\label{sec:transfer}

If the holographic bound (\ref{holographic}) can be
construed as telling us that a generic physical system in
4-D spacetime is fundamentally two-dimensional in space,
then a black hole in 4-D spacetime when viewed as an
information absorber or entropy emitter,  is fundamentally
one-dimensional in space
\cite{bek_mayo01}.  I proceed to explain.  

In the discipline treating information flow---communication 
theory---the notion of a channel is central.  In flat
spacetime a channel is a complete set of unidirectionally
propagating modes of some field, with the modes enumerated
by a single parameter.  For example, all electromagnetic
modes in free space with fixed wave vector {\it
direction\/} and particular linear polarization constitute
a channel; the modes are parametrized solely by frequency. 
An example would be a straight infinitely long coaxial
cable (which is well known to transmit all frequencies)
capped  at its entrance by the analog of a polaroid filter.
Acoustic and neutrino channels can also be defined.   Note
that a channel is intrinsically one-dimensional.  

What is the maximum {\it rate\/}, in quantum theory, at
which information may be transmitted through a channel for
prescribed power
$P$ ?   The answer has been known since the 1960's;
however, let me work with the particularly lucid version
given by Pendry
\cite{pendry}.  Pendry thinks of a possible signal state as
corresponding to a particular set of occupation numbers for
the various  propagating modes.  He assumes the channel is
uniform in the direction of propagation, which allows him
to label the modes by momentum $p$.  But he allows for
dispersion, so that a quantum with momentum $p$  has some
energy $\varepsilon(p)$. Then the propagation velocity of
the quanta is the group velocity
$\upsilon(p)=d\varepsilon(p)/dp$.  Up to a factor 
$\log_2 e$ the information rate capacity must equal the
maximal one-way entropy current for given $P$, which
obviously occurs for the thermal state, if one discards
from the latter the modes moving opposite the direction of
interest. 

Now the entropy $s(p) $ of any boson mode of momentum $p$ in
a thermal state (temperature $T$) is \cite{LL}
\begin{equation}  s(p)={\varepsilon(p)/T\over
e^{\varepsilon(p)/T}-1}-\ln \left(
1-e^{-\varepsilon(p)/T}\right),
\label{s}
\end{equation}  so the entropy current in one direction is
\begin{equation}
\dot S=\int^{\infty}_0 s(p)\thinspace\upsilon(p)\, {dp\over
2\pi\hbar}= \int^{\infty}_0 s(p)\thinspace
{d\varepsilon\over dp}\ {dp\over 2\pi\hbar},
\label{current}
\end{equation}  where $dp/2 \pi \hbar $ is the number of
modes per unit length in the interval $dp$ which propagate
in one direction.  This factor, when multiplied by the group
velocity, gives the one-way current of modes.   

Suppose $\varepsilon(p)$ is monotonic and extends over the
range
$[0,\infty)$; we may then cancel $dp$ and integrate over
$\varepsilon$.  Then after substitution of Eq.~(\ref{s})
and integration by parts we have
\begin{equation}
\dot S= {2\over T}\int^{\infty}_0 {\varepsilon\thinspace
\over e^{\varepsilon/T} -1}\ { d\varepsilon\over 2\pi\hbar}
={2\over T}\int^{\infty}_0 {\varepsilon(p)\over
e^{\varepsilon(p)/T} -1}\thinspace
\upsilon(p) \thinspace {dp\over 2\pi\hbar }. 
\label{entropy_flow}
\end{equation}  The first factor in each integrand is the
mean energy per mode, so that the integral represents the 
one-way power $P$ in the channel.  Thus 
\begin{equation}
\dot S=2P/T . 
\label{final}
\end{equation} The integral for $P$ in the first form of
Eq.~(\ref{entropy_flow}) can easily be done:
\begin{equation}  P = {\pi (T)^2\over 12 \hbar}.
\label{power}
\end{equation}  Eliminating $T$ between the last two
expressions gives Pendry's limit
\begin{equation}
\dot S=\left({\pi P\over 3\hbar}\right)^{1/2}\qquad {\rm
or}\qquad \dot I_{\rm max}= \left({\pi P\over
3\hbar}\right)^{1/2}\log_2 e.
\label{pendry_formula}
\end{equation}  For a fermion channel $P$ in
Eq.~(\ref{power}) is a factor 2 smaller, and consequently
$\dot S$ in Eq.~(\ref{pendry_formula}) is reduced by a
factor
$\surd 2$.
 
The function $\dot S(P)$ in Eq.~(\ref{pendry_formula}) is
the so called {\it capacity of a noiseless quantum
channel\/}.  Surprisingly, it is independent, not only of
the form of the mode velocity $\upsilon(p)$, but also of
its scale.  Thus the phonon channel capacity is as large as
the photon channel capacity despite the difference in
speeds. Why? Although phonons convey information at lower
speed, the energy of a phonon is proportionately smaller
than that of a photon in the equivalent mode.  Thus when
the capacities of channels harnessing various carriers are
expressed in terms of power, they turn out to involve the
same constants.   Formula.~(\ref{pendry_formula}) neatly
characterizes what we mean by one-dimensional transmission
of entropy or information.  It refers to transmission by
use of a single species of quantum and a specific
polarization; different species and alternative
polarizations engender separate channels.  Although framed
in a flat spacetime context, its lack of sensitivity to the
dispersion relation of the transmitting {\it milieu\/}
should make Pendry's limit relevant to curved spacetime
also.  This because electrodynamics in curved spacetime is
equivalent to flat spacetime electrodynamics in a suitable
dielectric and paramagnetic medium \cite{volkov}.  We shall
see in Sec.~\ref{sec:dump} that this hunch is justified. 

It is instructive to contrast the results just obtained
with the power and entropy emission rate in a single boson
polarization of a closed black body surface with
temperature $T$ and area $A$  in flat 4-D spacetime.  By
the Stefan-Boltzmann law this is
\begin{equation}  P={\pi^2 T^4 A\over 120
\hbar^3}\qquad\quad \dot S ={4\over 3}{P\over T}
\label{P}
\end{equation} 
 whereby
\begin{equation}
\dot S= {2\over 3}\left({2\pi^2AP^3\over
15\hbar^3}\right)^{1/4},
\label{3-D}
\end{equation}  [for fermions $P$ carries an extra factor
$7/8$ and formula (\ref{3-D}) an extra factor
$(8/7)^{1/4}$].   Our manifestly 3-D transmission system
deviates from the sleek formula (\ref{pendry_formula}) in
the exponent of $P$ and in the appearance of the measure
$A$ of the system.  In emission from a closed curve of
length $L$ in two-dimensional space, the factor $ (L
P^2)^{1/3}$ would replace $(A P^3)^{1/4}$.  We may thus
gather the dimensionality of the transmission system from
the exponent of $P$ in the expression $\dot S(P)$ [it is 
$n/(n+1)$ for
$D=n+1$ spacetime dimensions], as well as from the value of
the coefficient of $P/T$ in expressions for $\dot S$ like
(\ref{final}) or (\ref{P}) [it is
$(n+1)/n$].

Radiation from a Schwarzschild black hole in 4-D spacetime
is also given by Eqs.~(\ref{P}) (or their fermion version)
with $A=4\pi (2M)^2$ and $T=T_{\rm H}$, except we must
correct the expression for
$P$ by a species dependent factor $\bar\Gamma$ of order
unity
 \cite{page1}, and replace the $4/3$ in the expression for 
$\dot S$ by the species dependent factor $\nu$ already
mentioned in Sec.~\ref{sec:newbound}.  Eliminating $M$
between the equations we obtain, in lieu of Eq.~(\ref{3-D}),
\begin{equation}
\dot S = \left({\nu^2\bar\Gamma\pi P\over
480\hbar}\right)^{1/2}.
\label{BHlimit}
\end{equation}  (For fermions there is an extra factor $7/8$
inside the radical). This looks completely different from
the law (\ref{3-D}) for the hot closed surface because,
unlike for a hot body, a black hole's temperature is
related to its mass.

However, (\ref{BHlimit}) {\it is\/} of the same form as
Pendry's limit (\ref{pendry_formula}) for one-channel
transmission.  From Page \cite{page1,page2} we get
$\bar\Gamma=1.6267$ and  $\nu=1.5003$ for one photon
polarization, so the numerical coefficient of
(\ref{BHlimit}) is $15.1$ times that in
(\ref{pendry_formula}).  Repeating the above exercise for
one species of neutrinos we again find formulae like
(\ref{BHlimit}) and  (\ref{pendry_formula}), this time with
$\bar\Gamma=18.045$ and  $\nu=1.6391$;  the  numerical
coefficient of (\ref{BHlimit}) is $48.1$ times that of the
fermion version of (\ref{pendry_formula}). 

Thus when judged by its entropy emission properties, a
black hole in 4-D spacetime is more like a 1-D channel than
like a surface in 3-D space. Why is this ?  A formal answer
is that, because of the way $T_{\rm H}$ is related to the
black hole's radius $2M$,  Hawking emission prefers to
emerge in the lowest angular momentum mode possible.  To
exit with  impact parameter $<2M$ and angular momentum
$j\hbar$,  a quantum must have energy (momentum)
$\hbar\omega$ 
$>j\hbar/2M$.  But in the Hawking thermal distribution the
dominant $\hbar\omega$ is of order  $T_{\rm H}=\hbar(8\pi
M)^{-1}$.  Thus the emerging $j$'s tend to be small.  For
example, $97.9\%$ of the photon energy emerges in the $j=1$
modes ($j=0$ is forbidden for photons), and $96.3\%$ of the
neutrino power is in the
$j={\scriptstyle 1\over\scriptstyle 2}$ modes \cite{page1}. 
Thus the black hole emits as close to radially as
possible.  This means that, crudely speaking, it does so
through just one channel.

If a black hole emits entropy like a one-dimensional system,
we might guess it should absorb information like a one
dimensional system.  This hunch will be verified in
Sec.~\ref{sec:dump}.  As a first step I extend to curved
spacetime some of the insights regarding information flow. 

\section{Information Pulses in Curved Spacetime}
\label{sec:bursts}

The discussion in Sec.~\ref{sec:transfer} tacitly assumed
steady state streaming of information and energy.  But what
if information is delivered in pulses ?  Can one state a
bound generalizing (\ref{pendry_formula}) ? Can one include
effects of gravitation on the information transfer rate ? 
To answer these questions let us extend the notion of
channel to curved spacetime, at least to stationary curved
spacetime.  Again, a channel will be a complete set of
unidirectional modes of some field that can be enumerated
with a single parameter.  Each channel is characterized by
species of quanta, polarization (helicity), trajectory,
etc.  In Sec.~\ref{sec:transfer} I characterized the {\it
signal\/} in a particular channel by power.  For a pulse it
seems a better idea to use both the signal's duration
$\tau$ and its energy $E$.  Since in curved spacetime a
channel is not generally uniform, I choose to measure these
parameters in a local Lorentz frame (I shall show presently
that it does not matter which one).  With this precaution
sections of the channel may be treated as in flat spacetime.

How is the true $ I_{\rm max}$ of a pulse related to its $E$
and
$\tau$ ?  Since information is dimensionless, $I_{\rm max}$
must be a function of dimensionless combinations of $E$,
$\tau$, channel parameters and the fundamental constants
$c, \hbar$ and $G$:
\begin{equation}  I_{\rm max}=\Im (E\tau/\hbar,
GEc^{-5}\tau^{-1}). 
\label{imax}
\end{equation} Here $\Im(\xi,\varpi)$ is some nonnegative
valued function of the dimensionless parameters $\xi$ and
$\varpi$ characteristic of the  channel.  This is called
the ``characteristic information function'' (CIF)
\cite{bek88,bek_schiff}.  The shape of $\Im$ depend on
things like the polarization and nature of the transmitting
medium.  I shall assume this medium, if any, is
nondissipative and nondispersive.  Thus it is characterized
by a single signal velocity $c_s$; the dimensionless
parameter $c_s/c$ is one of the determinants of the shape
of $\Im$. I shall exclude channels which transmit massive
quanta, e.g. electrons, because rest mass is energy in a
form not useful for communication, so  that the strictest
limits on information flow should emerge for massless
signal carriers.  Hence masses do not enter into the shape
of $\Im$. The variable $\varpi\equiv GEc^{-5}\tau^{-1}$ is
of order of $E$ divided by the signal's self-potential
energy, and very large  for ordinary signals.  So I first
work with the limiting formula as
$\varpi\rightarrow\infty$.  

Let us check what happens in flat spacetime for steady state
signaling.  This implies we deal with a long stream of
information and that the signal can be characterized in a
suitable frame as statistically stationary.  The peak
information transfer {\it rate\/} and power can then be
inferred from a finite section of the signal of duration
$\tau$ bearing information $I_{\rm max}$ and energy $E$. It
should matter little how long a stretch in 
$\tau$ is  used so long as it is not too short, and $\dot
I_{\rm max}\equiv E\tau^{-1}$ should come out fully
determined by  the power
$P\equiv E\tau^{-1}$.  But this is consistent with
Eq.~(\ref{imax})  only if $\Im(\xi,\infty)\propto\surd
\xi$, for only then does $\tau$ cancel out.   With this
form we recover Pendry's formula  $\dot I_{\rm max}\propto
(P/\hbar)^{1/2}$, which we know to be the correct answer
for steady state flow in flat spacetime.  

The dividing line between steady state signaling and
signaling by  means of very long pulses is fuzzy. This
suggests that long pulse signals must also obey a Pendry
type formula, albeit approximately, c.f. \cite{marko}. The
law  $\dot I_{\rm max}\propto (P/\hbar)^{1/2}$ is evidently
inapplicable to {\it brief\/} information pulses.  For such
it may be replaced by a linear upper bound \cite{bek81b}
which may even transcend some of the limitations I imposed
to define $\Im(\xi,\varpi)$.  Consider the information $I$
to be encoded in some material structure ${\cal V}$ of
radius $R$ and rest energy $E$ which maintains its
integrity and dimensions as it travels from emitter to
receiver.  From Eq.~(\ref{original}) we have the strict
inequality $I<2\pi E R
\hbar^{-1}\log_2 e$.  The rate at which the information is
assimilated by the receiver is obviously restricted by the
local time
$\tau$ it takes for ${\cal V}$ to sweep by it.  From special
relativity
$\tau> 2R\gamma^{-1}$ with Lorentz's $\gamma$ accounting for
the Fitzgerald contraction of ${\cal V}$ in the frame of the
receiver.  Thus the peak information reception rate is
$I/\tau< \pi \gamma E \hbar^{-1}\log_2 e$, or
\begin{equation}
\dot I_{\rm rec}< \pi E_{\rm rec}\hbar^{-1}\log_2 e
\label{linear}
\end{equation}  where $E_{\rm rec}\equiv \gamma E$ is ${\cal
V}$'s energy as measured in the receiver's frame.  Bound
(\ref{linear}) replaces the information version of
Eq.~(\ref{pendry_formula}) when it comes to pulses.   Since
$\xi\equiv E_{\rm rec}\tau\hbar^{-1}$ we have the strict
linear bound $\Im(\xi,\infty)< (\pi \log_2 e)\xi $, a bound
which is supported by much evidence
\cite{bek88,bek_schiff} even when the signal has no rest
frame.  I must stress that the linear bound applies only
for small $\xi$; for $\xi\gg 1$ one may use Pendry's
formula. 

Detailed calculations \cite{bek_schiff,erice} show that
$E\tau$ is unchanged in the passage between Lorentz frames,
regardless of whether transmission is through a medium or
vacuum.  Thus the law $I_{\rm max}=\Im(E\tau/\hbar,\infty)$
is Lorentz invariant not only in vacuum where this is
required by relativity, but also in the presence of a
preferred frame established by the medium.  We can thus use
 $I_{\rm max}=\Im(E\tau/\hbar,\infty)$ both in the medium's
and in the signal emitter's (receiver's) Lorentz frame,
provided we do so at a fixed point. 

But how is the information transmission rate related at two
point along the channel ?   In flat spacetime, and in the
absence of dispersion, $E$ and $\tau$ are evidently
conserved with propagation.  And in the absence of
dissipation so is the information, so that  $I_{\rm
max}=\Im(E\tau/\hbar,\infty)$ is valid at every point along
the channel.  Once we are in stationary {\it curved\/}
spacetime, $E$ and $\tau$ are subject to redshift and
dilation effects, respectively.  However, the two effects
act in opposite senses so that $E\tau$ is again conserved
throughout the signal's flight.  Therefore, the formula is
meaningful throughout the channel.  In fact one can use
global values (as measured at infinity)  of
$E$ and $\tau$ in the formula.    In conclusion, one and
the same  formula limits information transmission,
propagation and reception rates.

When self-gravity of the signal pulse is not negligible,
$\varpi$ reappears as a possible argument of $\Im$. 
However, it is clear that $E/\tau$ is not a Lorentz scalar,
so inclusion of $\varpi$ would spoil the local Lorentz
invariance of Eq.~(\ref{imax}) and violate special
relativity for signals propagating in vacuum in a flat
background.  In a curved background there are further
arguments against inclusion of $\varpi$ in $\Im$.  In
vacuum we can use the requirement of local Lorentz
invariance to bar $\varpi$'s appearance, for a sufficiently
brief signal should admit being encompassed in its entirety
by local Lorentz frames.  Further, $\varpi$ evidently
decreases as the signal propagates {\it outward\/} in the
gravitational potential.  Thus,
$\Im(E\tau/\hbar,\varpi)$ would decrease either outwardly
(if $\Im$ increases with $\varpi$) or inwardly (if it
decreases as $\varpi$ increases).  If a signal's information
saturates the bound $\Im(E\tau/\hbar,\varpi)$ at some point
in the potential, then by conservation of information it
will exceed the bound once it has propagated somewhat in the
direction in which $\Im$ decreases.  This leads to a
contradiction. One could try to resolve the problem by
defining $I_{\rm max}$ only in terms of the {\it minimum\/}
value of $\varpi$ in the channel. But it seems strange that,
at least for brief signals, one cannot state $I_{\rm max}$
in terms of local quantities.  

Thus for signals propagating in vacuum in flat or curved
spacetime,
$\varpi$ cannot appear in $\Im$.  It is unclear whether this
conclusion extends to signal propagation in a medium.  For
one thing in curved spacetime a medium is never
homogeneous, which means, among other things, that $c_s$
varies.  This in itself puts in doubt our argument for
simplicity for the formula (\ref{imax}).

\section{Dumping Information into a Black Hole}
\label{sec:dump}
 
Suppose we are granted a certain power ${\cal P}$ to
accomplish the task of getting rid of a stream of possibly
compromising information by dumping it into a black hole.  
What is the maximum information dumping rate ?  

To answer this I first argue that if the signal comes from a
distance, it is transmitted down the hole through a single
channel---more or less---per field species and polarization.
Let us recall the rule for field mode counting.  In one
space dimension a length $L$ contains $(2\pi)^{-1}L\Delta k$
modes in the wave vector interval
$\Delta k$.  In 3-D we have $(2\pi)^{-3}L_x L_y L_z\Delta
k_x \Delta k_y \Delta k_z$ modes.  From this we may
conclude that if a flat 2-surface of area ${\cal A}$
radiates into a narrow solid angle $\Delta\Omega$  about its
normal, the number of modes out to a distance $L$ from it
whose wave vector magnitudes lie between
$k$ and $k+\Delta k$ is $(2\pi)^{-3}{\cal A}L k^2
\Delta\Omega
\Delta k$.  The factor $(2\pi)^{-1}L\Delta k$ is obviously
the number of modes emitted sequentially in each direction
and distinguished by their values of $k$. One can thus think
of ${\cal W}=(2\pi)^{-2}{\cal A} k^2 \Delta\Omega$ as the
number of active {\it channels\/}. 

Now let a transmitter with effective area ${\cal A}$ send an
information bearing signal towards a Schwarzschild black
hole of mass $M$ surrounded by vacuum and situated at
distance
$d\gg 2M$.  Let ${\cal A}$ be oriented with its normal
towards the black hole; evidently ${\cal A}<4\pi d^2$. As
viewed from the transmitter the black hole subtends solid
angle $\Delta\Omega=\pi(2M)^2/d^2$.   What should we take
for $k$ in the formula for ${\cal W}$ ?  Being interested
in the highest information for given energy (other things
being equal), we certainly want to use the smallest $k$
(smallest
$\hbar\omega$) possible.  But signals composed of too small
$k$'s will just be scattered by the black hole.  The
borderline is $k=2\pi/\lambda \approx 2\pi/(2M)$. With this
we find ${\cal W}< 4\pi^2$,  which means that, optimally,
information is transmitted down a black hole through just a
few channels per field species and polarization.  This is
independent of the scales $M$ and $d$ of the problem.  

In light of this we employ the one-channel formula
(\ref{imax}); according to our argument in
Sec.~\ref{sec:bursts}, we drop the argument $\varpi$. 
Further, since $E\tau$ is conserved in Schwarzschild
(stationary) spacetime, and closely equals ${\cal E} t$,
the values being measured at infinity, we have
$I_{\rm max}=\Im({\cal E} t/\hbar)$.  This for a pulse of
duration $t$ as seen from infinity.  If we are dealing with
a steady state stream of energy and information
($t\rightarrow\infty$ and ${\cal E}\rightarrow\infty$ with
${\cal P}\equiv
\lim ({\cal E}/t)$ finite), we have, by the logic of the
paragraph following Eq.~(\ref{imax}), that the maximum
information disposal rate into the black hole is $\dot
I_{\rm max}\sim ({\cal P}/\hbar)^{1/2}$, as hinted at the
end of Sec.~\ref{sec:transfer}.  We thus discover that the
power required to dispose of information into a black hole
grows {\it quadratically\/} with the information dumping
rate.    

{\bf Acknowledgments\hskip 6pt} This research is supported
by grant No. 129/00-1 of the Israel Science Foundation.

\end{document}